\documentclass{emulateapj}

\def\arcs{$''$}

\def\hub{\ifmmode H_\circ\else H$_\circ$\fi}

\shorttitle{Old Star Clusters in M31} \shortauthors{Ma et al.}

\begin{document}
\slugcomment{AJ, in press}
\title{SPECTRAL ENERGY DISTRIBUTIONS AND MASSES of 304 M31 OLD STAR CLUSTERS}

\author{
Jun Ma\altaffilmark{1}, Song Wang\altaffilmark{1,2}, Zhenyu Wu\altaffilmark{1}, Tianmeng Zhang\altaffilmark{1}, Hu Zou\altaffilmark{1}, Jun dan, Nie\altaffilmark{1}, Zhiming Zhou\altaffilmark{1}, Xu Zhou\altaffilmark{1}, Jianghua Wu\altaffilmark{3}, Cuihua Du\altaffilmark{4}, Qirong Yuan\altaffilmark{5}}

\altaffiltext{1}{Key Laboratory of Optical Astronomy, National Astronomical Observatories, Chinese Academy of Sciences, Beijing 100012, China; majun@nao.cas.cn}

\altaffiltext{2}{University of Chinese Academy of Sciences, Beijing 100039, China}

\altaffiltext{3}{Department of Astronomy, Beijing Normal University, Beijing 100875, China}

\altaffiltext{4}{College of Physical Sciences, Graduate University of the Chinese Academy of Sciences, Beijing 100049, China}

\altaffiltext{5}{Department of Physics, Nanjing Normal University, WenYuan Road 1, Nanjing 210046, China}

\email{}

\begin{abstract}
This paper presents CCD multicolor photometry for 304 old star clusters in the nearby spiral galaxy M31. Of which photometry of 55 star clusters is first obtained. The observations were carried out as a part of the Beijing--Arizona--Taiwan--Connecticut (BATC) Multicolor Sky Survey from 1995 February to 2008 March, using 15 intermediate-band filters covering 3000--10000 \AA. Detailed comparisons show that our photometry is in agreement with previous measurements. Based on the ages and metallicities from Caldwell et al. and the photometric measurements here, we estimated the clusters' masses by comparing their multicolor photometry with stellar population synthesis models. The results show that the sample clusters have masses between $\sim 3\times10^4 M_\odot$ and $\sim 10^7 M_\odot$ with the peak of $\sim 4\times10^5 M_\odot$. The masses here are in good agreement with those in previous studies. Combined with the masses of young star clusters of M31 from Wang et al., we find that the peak of mass of old clusters is ten times that of young clusters.
\end{abstract}

\keywords{catalogs -- galaxies: individual (M31) -- galaxies: spiral -- galaxies: star clusters: general}

\section{INTRODUCTION}
\label{s:intro}

The study of star clusters is important for understanding the formation and evolution of galaxies. Individual clusters, which encapsulate at least a partial history of the parent galaxy's evolution, can provide a unique  laboratory for studying the ongoing and past star formation in the parent galaxy, since significant star cluster formation is typically produced by major star-forming episodes in a galaxy.

M31 (NGC 224), the largest galaxy in the Local Group, is located at a distance of $\sim770$ kpc from us \citep[see][and references therein]{cald11}. It is so near and has a great number of star clusters that M31 offers us an excellent environment for detailed, more accurate investigations of a star cluster system.

\citet{hubble32} did the pioneering work of M31 star clusters, in which he presented photographic magnitudes for 140 globular cluster (GC) candidates. Then, a number of catalogs of M31 GC candidates were published. Later, a large survey of \citet{sarg77} presented discoveries of many new GCs well out in the M31 halo, and these authors also completely collected and revised all previous lists of GC candidates. Their final catalog includes 355 GC candidates based on a uniform selection scale. Another three large surveys of M31 GC candidates were performed by \citet{Crampton85}, \citet{batt87}, and \citet{kim07}. Spectroscopic study of M31 GCs was begun by \citet{Sidney69}, followed by a significant number of authors. Of which four large spectroscopic studies were given by \citet{hbk91}, \citet{per02}, \citet{lee08}, and \citet{cald09}. The first comprehensive catalog including photometric and spectroscopic data for M31 GCs was assembled by \citet{bh00}, who originally constructed a single master catalog incorporating the photometric and spectroscopic data in all of the individual catalogs before 2000 including their new photometric and spectroscopic data. This catalog contains 435 clusters and cluster candidates. Based on this master catalog, these authors firstly presented the integrated photometric and spectroscopic properties of the globular cluster system of M31. The Revised Bologna Catalog (RBC) of M31 GCs was published by \citet{gall04} and since then it has been revised including all of the newest published important data. RBC is the most extensive and commonly used. \citet{cald09} presented a new catalog containing 670 likely star clusters in M31, all with updated high-quality coordinates accurate to $0.2''$, based on the images from either the Local Group Galaxies Survey (hereafter LGGS) \citep{massey06} or the Digitized Sky Survey (DSS). These authors derived high-quality spectra for $\sim 1000$ star clusters and star cluster candidates in M31 taken with the Hectospec spectrograph on the 6.5 m MMT. Based on the archival {\it Hubble Space Telescope} ({\it HST}) images, and images from the LGGS, and the spectra they observed, these authors confirmed cluster classifications. In addition, \citet{cald09} distinguished M31's young clusters from old, and estimated ages, reddening values, and masses for 140 young clusters by comparing the observed spectra with the simple stellar population (SSP) models. Subsequently, \citet{cald11} determined the mean metallicities, ages and reddening values for 323 M31 old star clusters, and presented some properties of them such as the metallicity distribution. \citet{wang12} determined photometry in 15 intermediate-band filters of the BATC system for 135 young star clusters in M31 from \citet{cald09} and \citet{cald11}, and estimated their ages and masses by comparing their spectral energy distributions (SEDs) with SSP models, and presented integrated properties for these young star clusters.

The distance modulus was taken to be $(m-M)_0=24.43$ throughout as \citet{cald11} adopted.

In this paper, we perform aperture photometry in 15 intermediate-band filters of the BATC system for 304 old star clusters in M31 from \citet{cald11}. By comparing the observed multicolor photometry with the SSP models, we derive their masses. This paper is organized as follows. Section 2 describes the sample star clusters and presents the BATC observations. In Section 3, we estimate masses of the sample clusters. Lastly, our discussions and conclusions are presented in Sections 4 and 5.

\section{SAMPLE OF STAR CLUSTERS , OBSERVATIONS, AND DATA REDUCTION}
\label{s:data}

\subsection{Sample of Star Clusters}
\label{s:samp}

The sample of star clusters in this paper is from \citet{cald11}, who determined the mean metallicities, ages and reddening values for 323 old star clusters in M31 based on the spectroscopic data obtained with the Hectospec fiber positioner and spectrograph on the 6.5 m MMT. Except for five star clusters, all the other sample clusters were observed with 15 intermediate-band filters of the BATC photometric system. However, there are 13 clusters for which we cannot present their accurate photometry with different following reasons: (1) some clusters are very close to other objects, (2) some clusters are very faint and the signal-to-noise ratio (S/N) is low, (3) some clusters are superimposed onto a very uneven background. In addition, B065-G126 falls in the bleeding CCD column of a saturated star, of which we cannot present its photometry. Thus, we will perform photometry for the remaining 304 clusters in the BATC photometric system.

\subsection{Archival Images of the BATC Sky Survey for M31 Field}

The M31 field is part of a galaxy calibration program of the BATC Multicolor Sky Survey. The BATC program uses the 60/90 cm Schmidt Telescope at the Xinglong Station of the National Astronomical Observatories, Chinese Academy of Sciences (NAOC). This system includes 15 intermediate-band filters, covering a range of wavelength from 3000 to 10000 \AA~\citep[see][for details]{fan96}. Before February 2006, a Ford Aerospace $2{\rm k}\times2{\rm k}$ thick CCD camera was applied, which has a pixel size of 15 $\mu\rm{m}$ and a field of view of $58^{\prime} \times 58^{\prime}$, resulting in a resolution of $1''.67~\rm{pixel}^{-1}$. After February 2006, a new $4{\rm k}\times4{\rm k}$ CCD with a pixel size of 12 $\mu$m was used with a resolution of $1''.36$ pixel$^{-1}$ \citep{fan09}. We obtained 143.9 hours of imaging of the M31 field covering about 6 square degrees, consisting of 447 images, through the set of 15 filters in five observing runs from 1995 to 2008, spanning 13 years \citep[see][for details]{fan09}.

Figure 1 shows the spatial distribution of 304 sample clusters (black dots) and the M31 fields observed with the BATC multicolor system, in which a box only indicates a field view of $58^{\prime}$ $\times $ $58^{\prime}$ of the thick CCD camera.

\begin{figure*}
\centerline{\includegraphics[scale=0.9,angle=0]{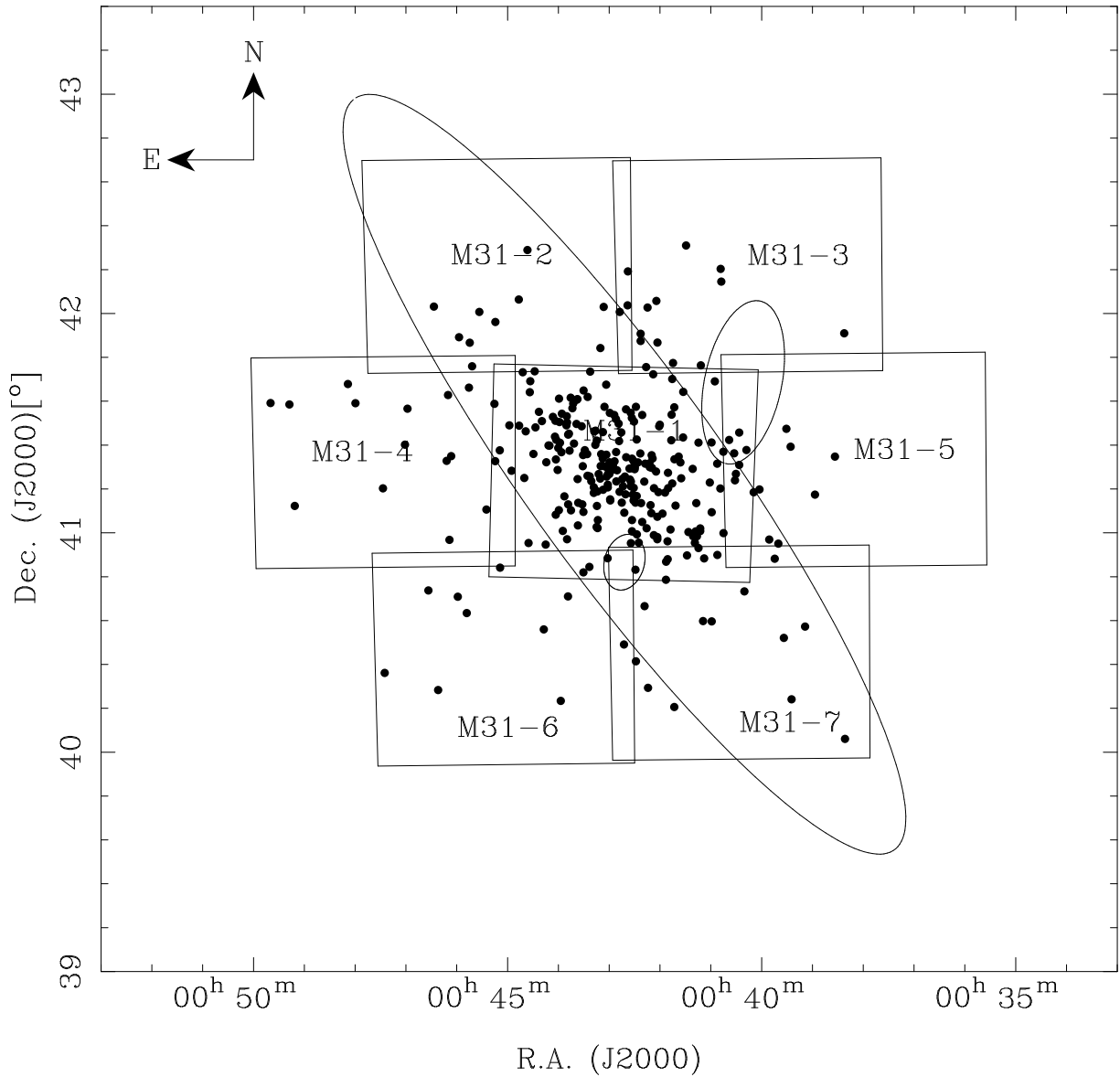}}
\caption[]{Spatial distribution of 304 sample clusters of M31 indicated by black dots. A box represents a field view of $58^{\prime}$ $\times $ $58^{\prime}$. The large ellipse is the M31 disk/halo boundary as defined by \citet{rac91}; the two small ellipses are the $D_{25}$ isophotes of NGC 205 (northwest) and M32 (southeast).}
\end{figure*}

\subsection{Integrated Photometry}
\label{s:phot}

We performed aperture photometry of 304 sample clusters found in the BATC images in all of the 15 intermediate-band filters to provide a comprehensive and homogeneous photometric catalog for them. Of which photometry of 55 star clusters is first obtained. We want to point out that, \citet{jiang03}, \citet{ma06,ma09b}, \citet{fan09}, and \citet{wang10} had presented photometry for 249 sample star clusters in the BATC photometric system, however, we also perform photometry for them in order to present a comprehensive and homogeneous photometric catalog. In addition, \citet{jiang03} and \citet{ma06,ma09b} did not present photometry in the $a$ and $b$ bands for 184 sample clusters, since the images including these star clusters in these two filters had not been observed at that time. The photometric routine we used is {\sc iraf/daophot} \citep{stet87}. To determine the total luminosity of each cluster, we produced curve of growth from $g$-band photometry obtained through apertures with radii in the range 2--11 pixel with 1 pixel increments. These were used to determine the aperture size required to enclose the total cluster light. The most appropriate photometric radius that includes all light from the objects, but excludes (as much as possible and to the extent that was obvious) extraneous sources is adopted independently for each cluster. This method ensures that we can determine the total cluster luminosity correctly, especially the use of small apertures for small clusters can maximize the S/N and minimize the contamination from nearby sources. In addition, we have checked the aperture of every sample star cluster considered here by visual examination to make sure that it was large enough to include all light from this object, but not so large as to be contaminated by other sources. The local sky background was measured in an annulus with an inner radius which being larger 1 pixel than photometric radius and $\sim 8.4''$  wide, in which the mode was used. Table 1 lists our new magnitudes and the aperture radii used, with errors given by {\sc iraf/daophot}. Column (1) gives the cluster names, which follow the naming convention of \citet{cald11}. Columns (2) to (16) present the magnitudes in the 15 BATC intermediate-band filters. The $1\sigma$ magnitude uncertainties from {\sc DAOPHOT} are listed for each object on the second line for the corresponding bands. Column (17) is the photometric aperture adopted here. Some sample clusters fall in the edges of the images; B124-NB10 is saturated in the $i$-filter image; B156-G211 and B515 fall in the bleeding CCD column of a saturated star in the $i$-filter image; There is a bright source near B231-G285 in the $b$-filter image. In these cases, we cannot present star clusters' photometry. In addition, photometry of some sample star clusters in some bands (mainly in the $a$ and $b$ bands) cannot be presented here because of low S/Ns. Figure 2 shows curve of growth from different bands from $a$ to $p$ for 4 representative star clusters selected randomly: one compact and bright, B097-G159, one extended and faint, B199-G248, one compact and faint, B200, and one extended and bright, B225-G280. In Figure 2, the most appropriate photometric radius adopted here is indicated by triangles. For B199-G248 and B225-G280, the most appropriate photometric radius obtained from the $g$ band is also most fitted in the other bands. For B097-G159, curve of growth from the $b$ band shows that the most appropriate photometric radius is $5.01''$, and the magnitude difference between $5.01''$ and $6.68''$ is 0.058 mag. However, the S/Ns of B097-G159 in all the bands are low, and the photometric errors are larger than 0.1 mag in all the photometric apertures. So, when we consider the photometric errors, the photometric radius of $6.68''$ is also appropriate in the $b$ band. For B200, the S/Ns in all the bands are very low. When we consider the photometric errors, the most appropriate photometric radius obtained from the $g$ band is also most fitted in the other bands.

Figure 3 presents the correlation between the radius of aperture adopted here and the magnitude in the $g$ band for the sample star clusters. From Figure 3, we can see that, in general, the brighter a star cluster, the larger the aperture adopted here, although the scatter of luminosity is large for the same size of aperture.

\begin{figure*}
\centerline{\includegraphics[scale=0.65,angle=-90]{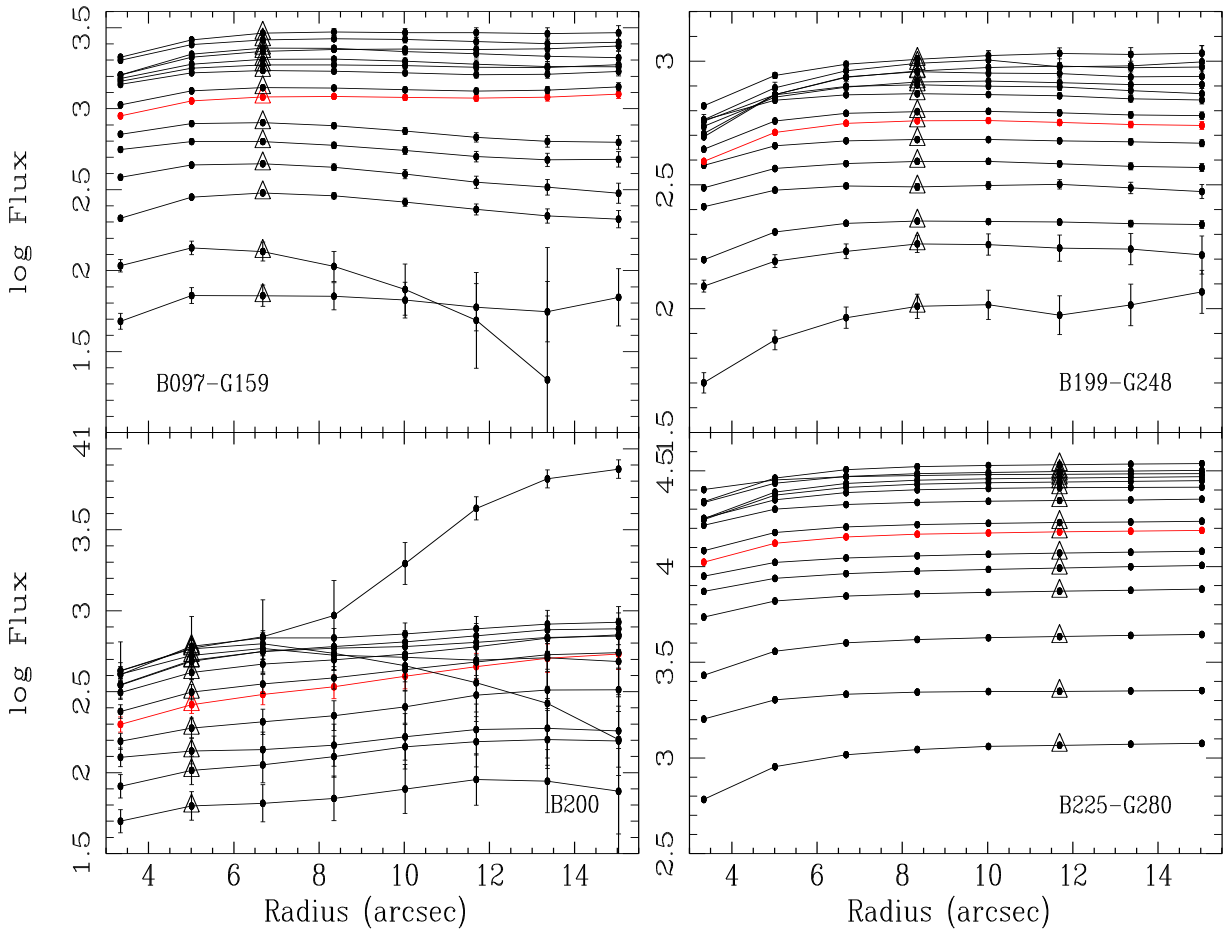}}
\caption[]{Curve of growth for different bands from $a$ to $p$ for four GCs of M31: one compact and bright, B097-G159, one extended and faint, B199-G248, one compact and faint, B200, one extended and bright, B225-G280, which are selected randomly. Triangles indicate the radii of the apertures adopted for photometry. From bottom to top are those of $a-p$. Curve of growth in the $g$ band is indicted in red.}
\end{figure*}

\begin{figure*}
\centerline{
\includegraphics[scale=0.65,angle=-90]{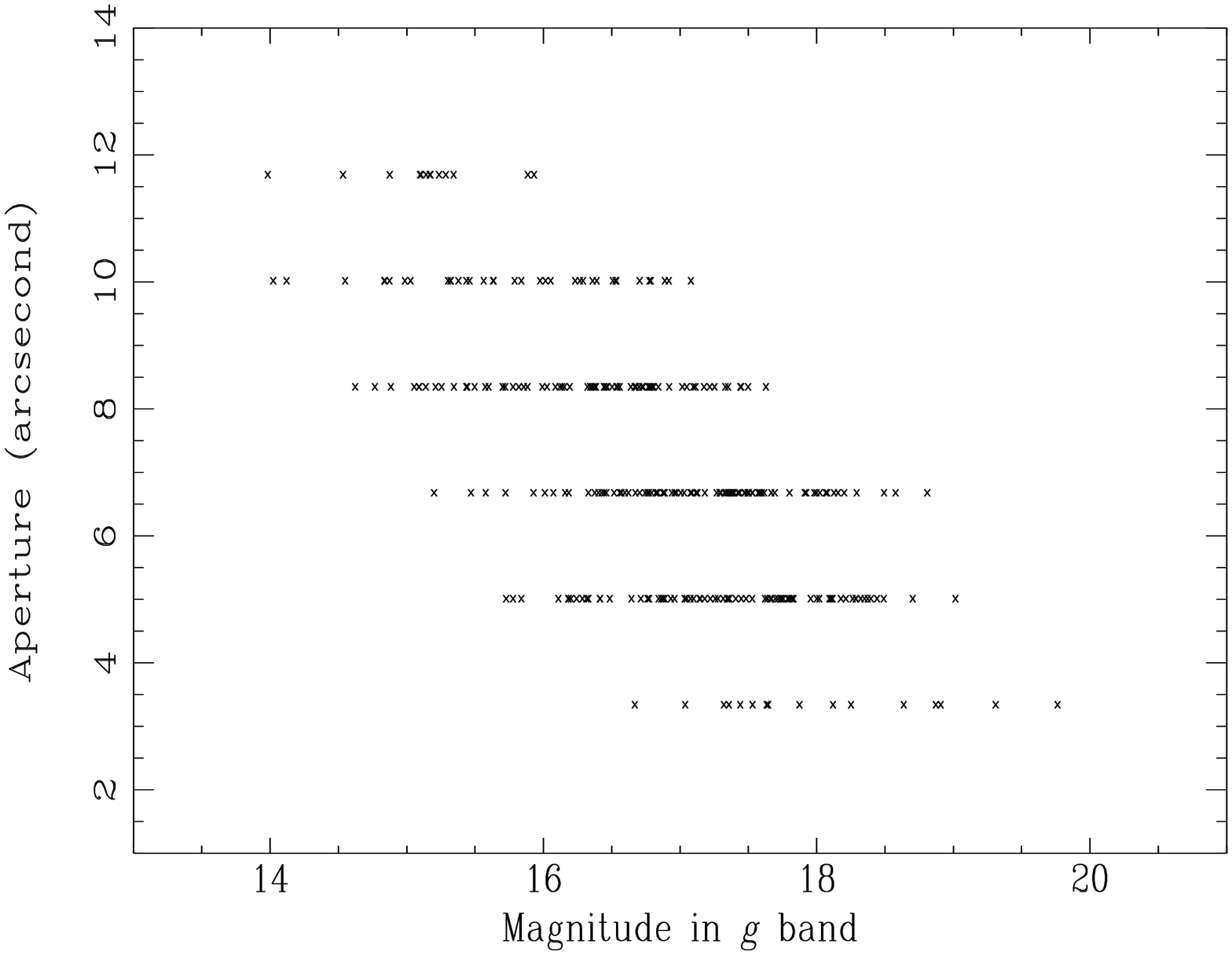}}
\caption[]{Correlation between the sizes of aperture adopted here and the magnitudes obtained here for star clusters.}
\end{figure*}

\subsection{Comparison with Previous Photometry in the $a-p$ Bands}

In literature, \citet{jiang03} and \citet{ma06,ma09b} presented photometry of 184 sample clusters in 13 intermediate-band filters, and \citet{fan09} and \citet{wang10} presented photometry of 65 sample clusters in 15 intermediate-band filters, respectively. These authors adopted the same aperture diameter of $10.02''$ for all their star clusters when performing photometry, and aperture corrections are computed using isolated stars. Figures 4--6 compare our new photometry with that of previous studies \citep{jiang03,ma06,ma09b,fan09,wang10}, which show that our new photometry is in good agreement with that of previous studies for bright objects, while the photometric measurements published by previous studies seem to be somewhat brighter than ours for faint clusters. In addition, there are 3 star clusters (AU010, B104-NB5, and B126-G184), of which the magnitude scatters in some bands between this study and previous studies are larger than 1.0, i.e., our new magnitudes are fainter than those of previous studies \citep{jiang03,ma06,ma09b,fan09,wang10}. We reported their names in Figures 4--6 (black circles). We checked the images of these three star clusters observed by the BATC system, and found that they are all compact and faint, and near the center of M31. In addition, they are superimposed onto an uneven background which is also bright. Photometric aperture radii of them adopted here are $3.34''$, and larger aperture sizes would cause a large photometric scatter.

\begin{figure*}
\centerline{
\includegraphics[height=175mm,angle=-90]{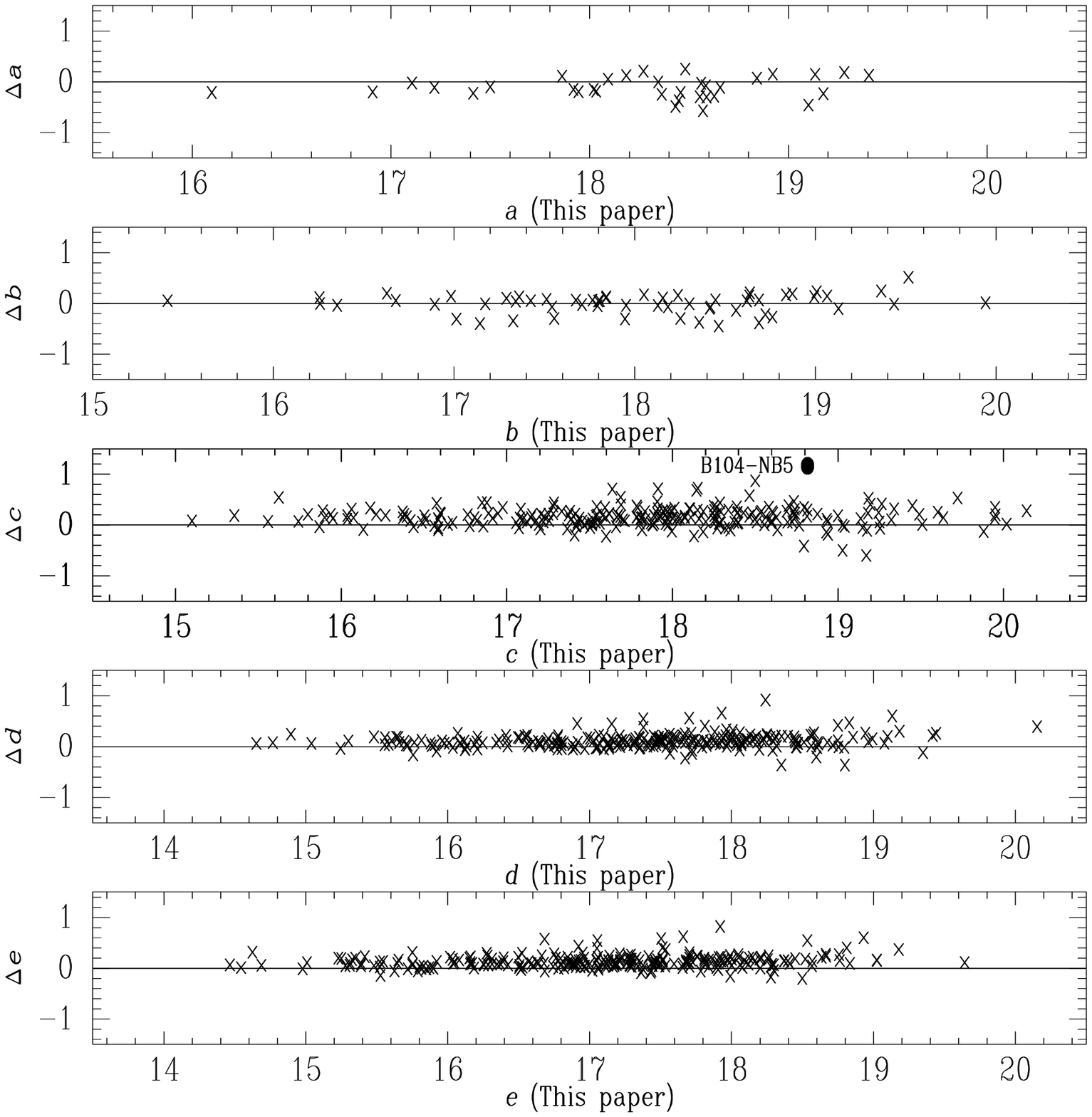}}
\vspace{-0.0cm}
\caption{Comparisons of our photometry of M31 star clusters in the $a-e$ bands with that of previous studies \citep{jiang03,ma06,ma09b,fan09,wang10}. The photometric offsets and rms scatters of the differences between their data and our new magnitudes are: $\Delta a = -0.112\pm0.036$ with $\sigma=0.209$ (34 clusters in common), $\Delta b = -0.006\pm0.025$ with $\sigma=0.189$ (59 clusters in common), $\Delta c = 0.164\pm0.013$ with $\sigma=0.199$ (248 clusters in common), $\Delta d = 0.113\pm0.009$ with $\sigma=0.141$ (248 clusters in common), and $\Delta e = 0.132\pm0.008$ with $\sigma=0.129$ (249 clusters in common) (this study minus previous studies).}
\end{figure*}

\begin{figure*}
\centerline{
\includegraphics[height=175mm,angle=-90]{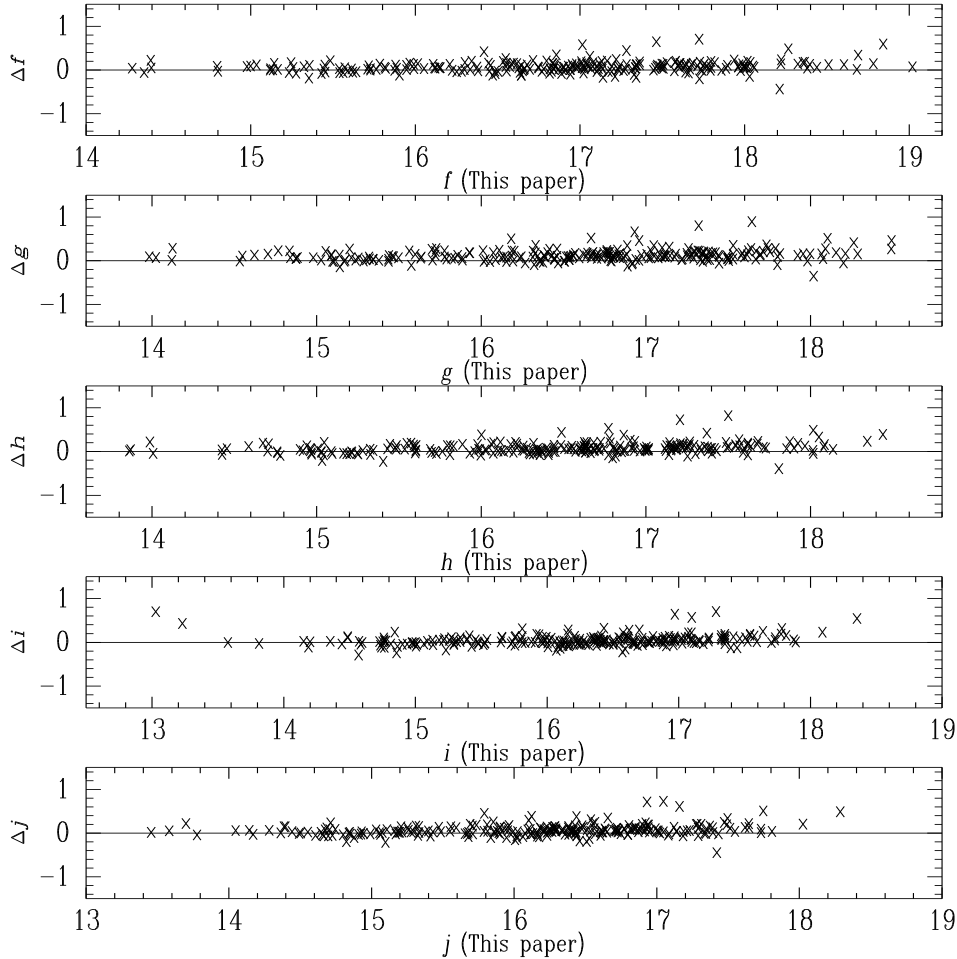}}
\vspace{-0.0cm}
\caption{Comparisons of our photometry of M31 star clusters in the $f-j$ bands with that of previous studies \citep{jiang03,ma06,ma09b,fan09,wang10}. The photometric offsets and rms scatters of the differences between their data and our new magnitudes are: $\Delta f = 0.074\pm0.008$ with $\sigma=0.127$ (249 clusters in common), $\Delta g = 0.126\pm0.009$ with $\sigma=0.135$ (249 clusters in common), $\Delta h = 0.079\pm0.008$ with $\sigma=0.127$ (249 clusters in common), $\Delta i = 0.049\pm0.008$ with $\sigma=0.127$ (247 clusters in common), and $\Delta j = 0.069\pm0.008$ with $\sigma=0.129$ (249 clusters in common) (this study minus previous studies).}
\end{figure*}

\begin{figure*}
\centerline{
\includegraphics[height=175mm,angle=-90]{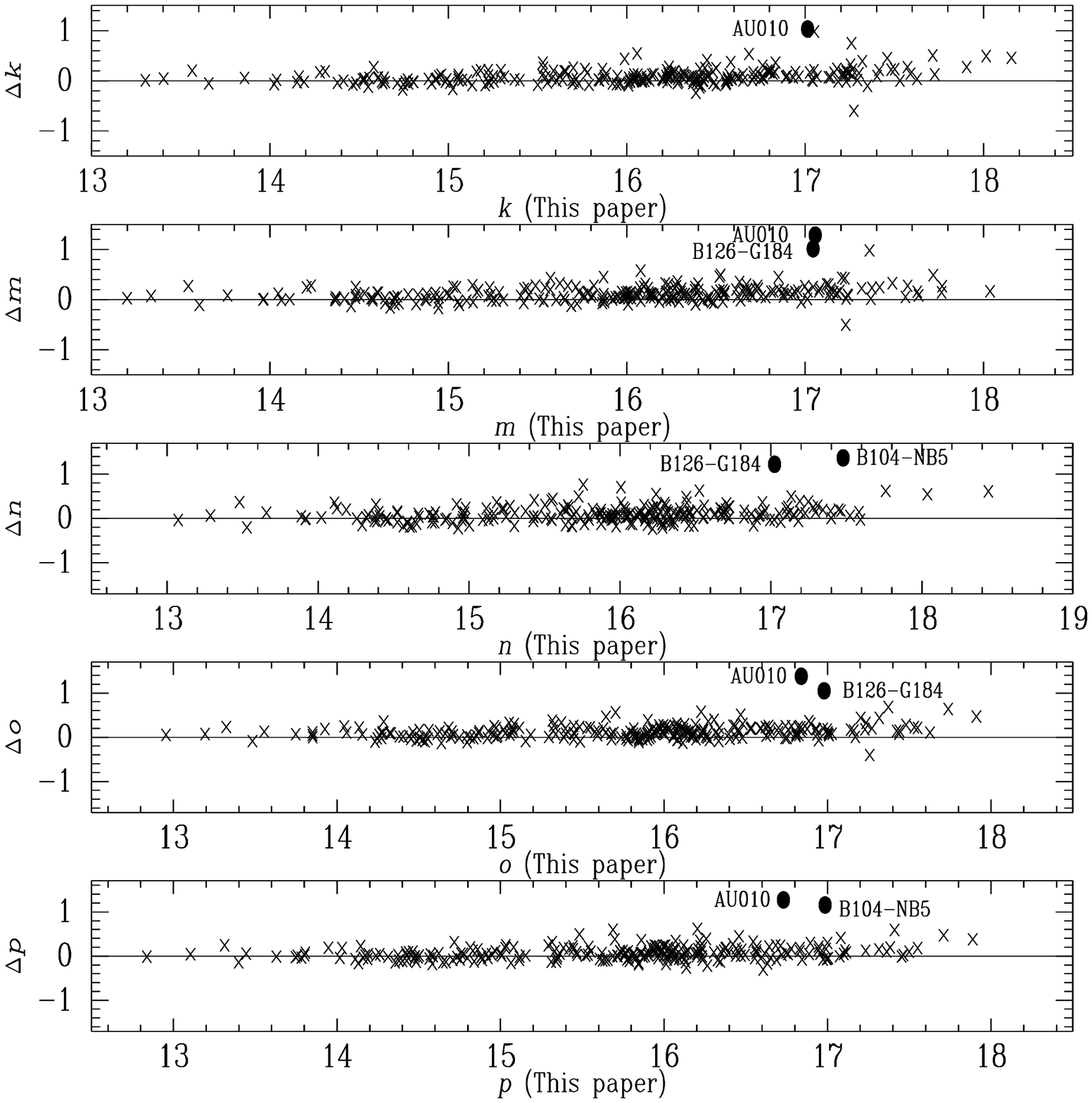}}
\vspace{-0.0cm}
\caption{Comparisons of our photometry of M31 star clusters in the $k-p$ bands with that of previous studies \citep{jiang03,ma06,ma09b,fan09,wang10}. The photometric offsets and rms scatters of the differences between their data and our new magnitudes are: $\Delta k = 0.089\pm0.010$ with $\sigma=0.163$ (249 clusters in common), $\Delta m = 0.123\pm0.011$ with $\sigma=0.172$ (249 clusters in common), $\Delta n = 0.109\pm0.013$ with $\sigma=0.204$ (244 clusters in common), $\Delta o = 0.130\pm0.011$ with $\sigma=0.171$ (249 clusters in common), and $\Delta p = 0.077\pm0.011$ with $\sigma=0.179$ (247 clusters in common) (this study minus previous studies).}
\end{figure*}

\subsection{Comparison with Previous Photometry in the $UBVRI$ Bands}

To examine the quality and reliability of our photometry, we compared the aperture magnitudes of the sample star clusters considered here with the magnitudes collected from various sources in the latest Revised Bologna Catalog of M31 GCs and candidates (RBC V.5) \citep{gall04,gall06,gall07,gall09}, and with previous measurements by \citet{bh00}, \citet{fan10}, and \citet{Peacock10}. First, we transformed the BATC intermediate-band system to the broad-band system using the relationships between these two systems derived by \citet{zhou03}:

\begin{equation}
m_U=m_{b}+0.6801(m_{a}-m_{b})-0.8982\pm0.143, \quad
\end{equation}

\begin{equation}
m_B=m_{d}+0.2201(m_{c}-m_{e})+0.1278\pm0.076, \quad
\end{equation}

\begin{equation}
m_V=m_{g}+0.3292(m_{f}-m_{h})+0.0476\pm0.027, \quad
\end{equation}

\begin{equation}
m_R=m_{i}+0.1036\pm0.055, \quad
\end{equation}

\begin{equation}
m_I=m_{o}+0.7190(m_{n}-m_{p})-0.2994\pm0.064.
\end{equation}

\begin{figure*}
\centerline{
\includegraphics[height=175mm,angle=-90]{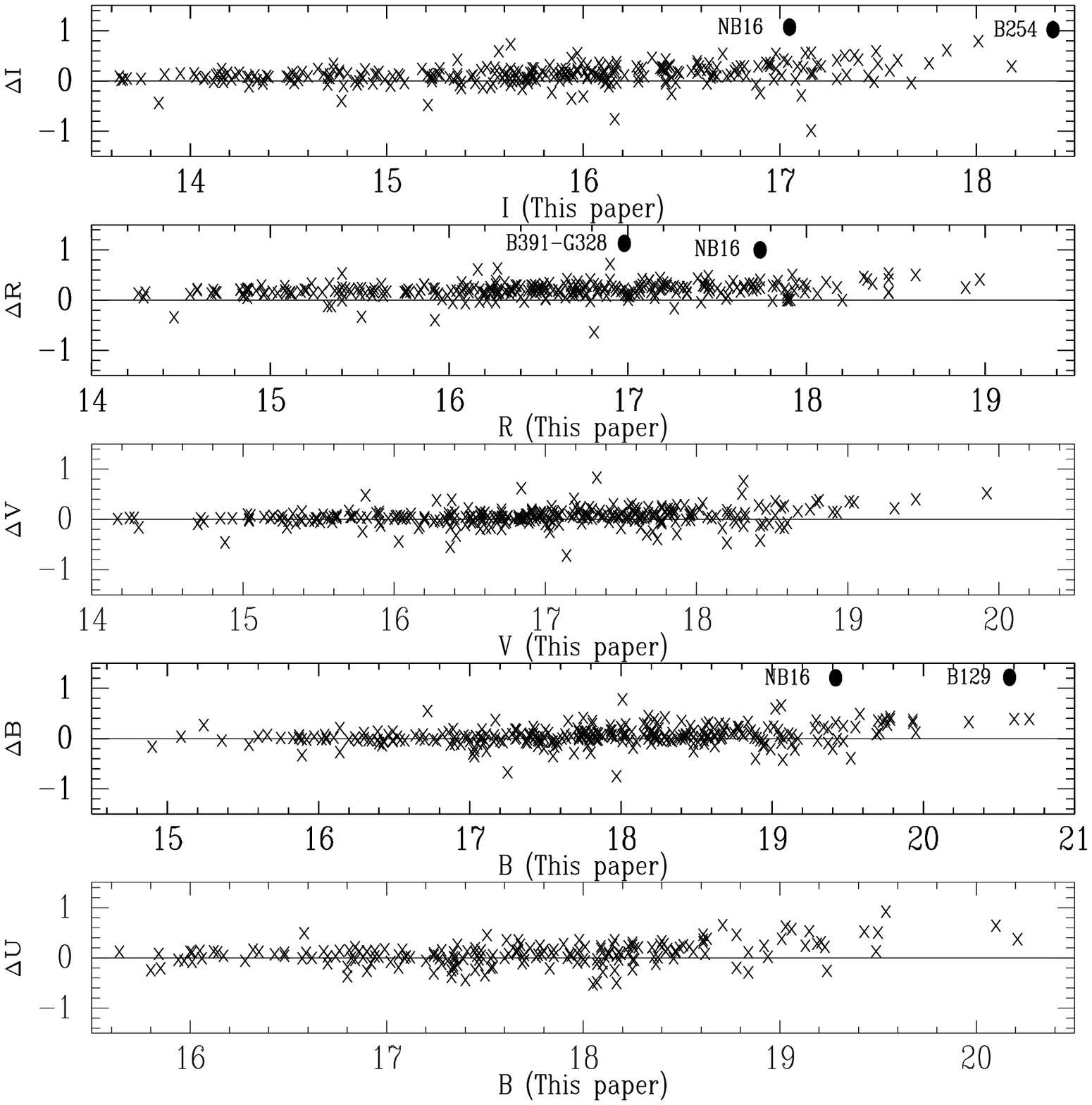}}
\vspace{-0.0cm}
\caption{Comparisons of our photometry of M31 star clusters in the $UBVRI$ bands with previous photometric data collected in \citet{{gall04}}. The photometric offsets and rms scatters of the differences between their data and our new magnitudes are: $\Delta U = 0.073\pm0.016$ with $\sigma=0.219$ (194 clusters in common), $\Delta B = 0.064\pm0.012$ with $\sigma=0.205$ (301 clusters in common), $\Delta V = 0.056\pm0.010$ with $\sigma=0.169$ (301 clusters in common), $\Delta R = 0.204\pm0.009$ with $\sigma=0.161$ (298 clusters in common), and $\Delta I = 0.136\pm0.012$ with $\sigma=0.210$ (284 clusters in common) (this study minus Galleti et al. 2004).}
\end{figure*}

\begin{figure*}
\centerline{
\includegraphics[height=175mm,angle=-90]{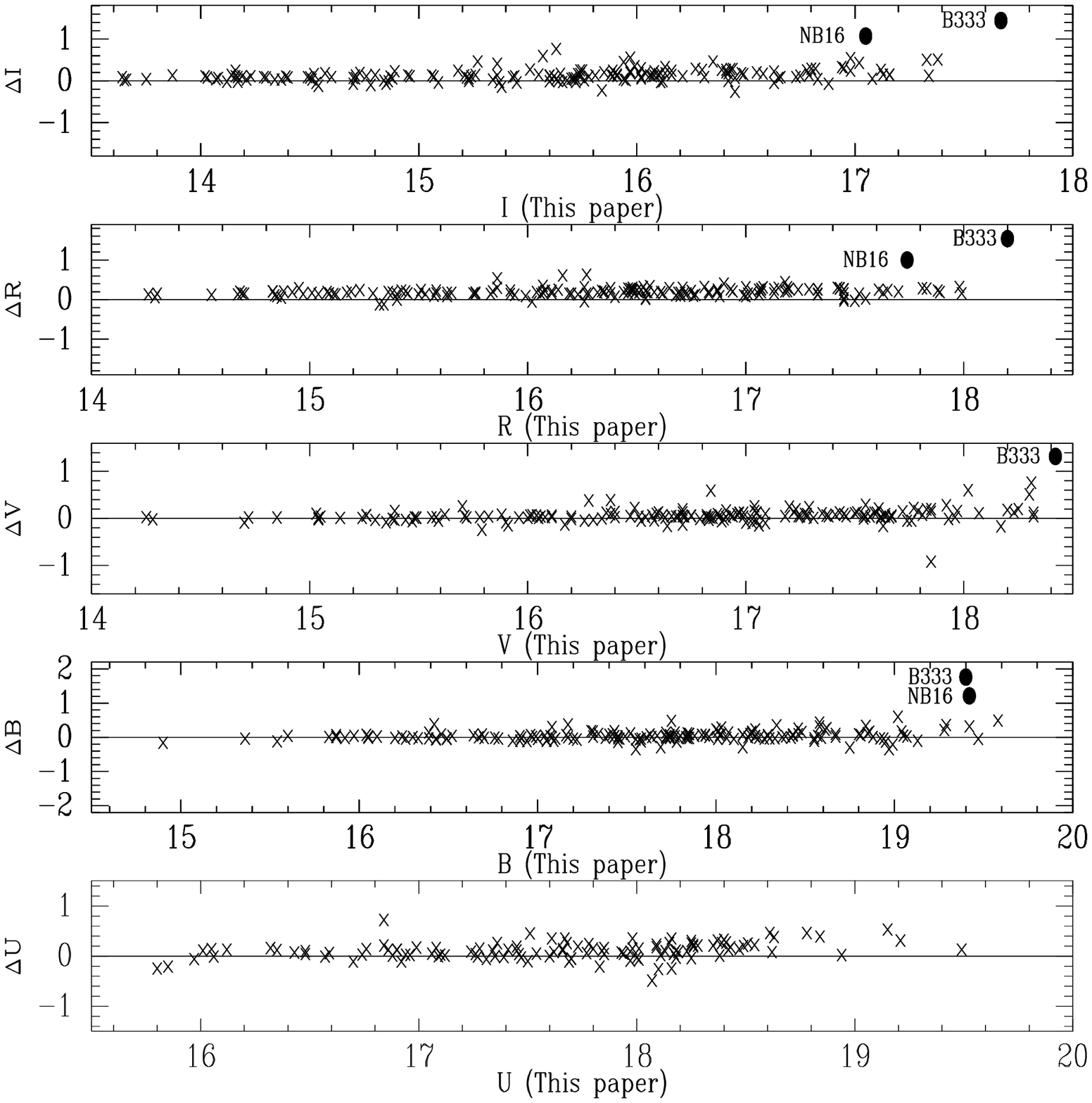}}
\vspace{0.0cm}
\caption{Comparisons of our photometry of M31 star clusters in the $UBVRI$ bands with that of \citet{bh00}. The photometric offsets and rms scatters of the differences between their measurements and our new magnitudes are: $\Delta U = 0.107\pm0.016$ with $\sigma=0.172$ (119 clusters in common), $\Delta B = 0.059\pm0.016$ with $\sigma=0.211$ (174 clusters in common), $\Delta V = 0.068\pm0.013$ with $\sigma=0.177$ (174 clusters in common), $\Delta R = 0.204\pm0.012$ with $\sigma=0.157$ (172 clusters in common), and $\Delta I = 0.147\pm0.014$ with $\sigma=0.189$ (173 clusters in common) (this study minus Barmby et al. 2000).}
\end{figure*}

\begin{figure*}
\centerline{
\includegraphics[height=175mm,angle=-90]{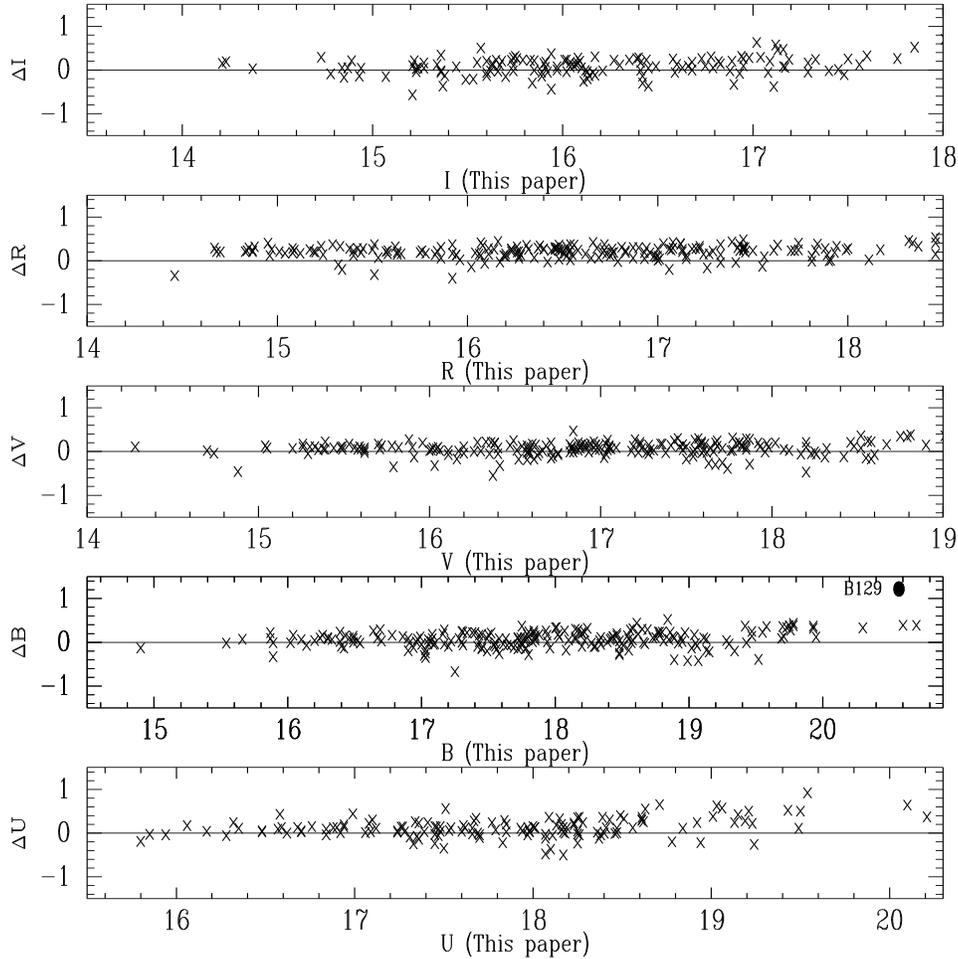}}
\vspace{0.0cm}
\caption{Comparisons of our photometry of M31 star clusters in the $UBVRI$ bands with that of \citet{fan10}. The photometric offsets and rms scatters of the differences between their measurements and our new magnitudes are: $\Delta U = 0.121\pm0.019$ with $\sigma=0.221$ (143 clusters in common), $\Delta B = 0.071\pm0.014$ with $\sigma=0.200$ (215 clusters in common), $\Delta V = 0.070\pm0.011$ with $\sigma=0.159$  (216 clusters in common), $\Delta R = 0.206\pm0.010$ with $\sigma=0.146$ (208 clusters in common), and $\Delta I = 0.093\pm0.018$ with $\sigma=0.218$ (142 clusters in common) (this study minus Fan et al. 2010).}
\end{figure*}

\begin{figure*}
\centerline{
\includegraphics[height=175mm,angle=-90]{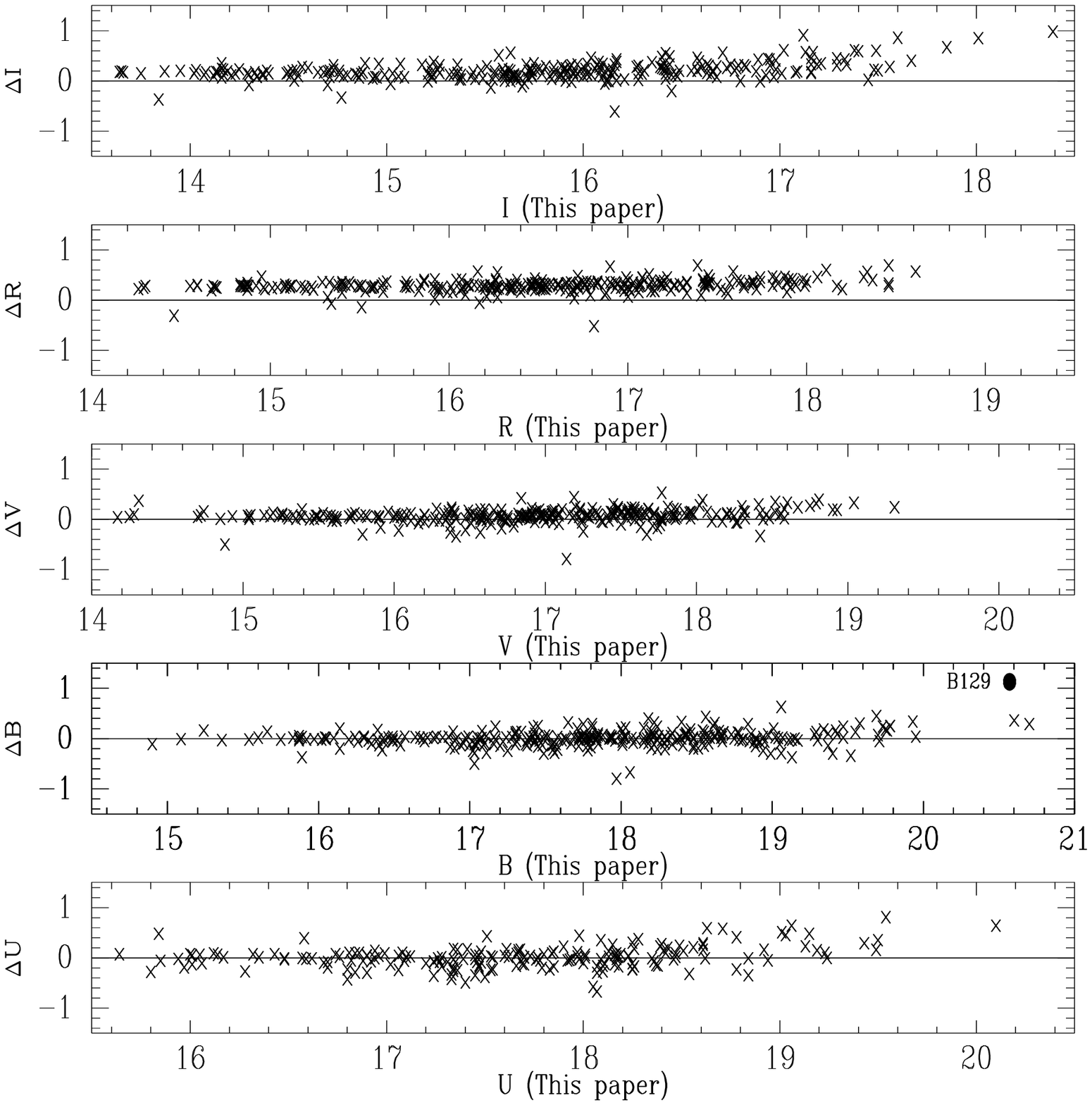}}
\vspace{0.0cm}
\caption{Comparisons of our photometry of M31 star clusters in the $UBVRI$ bands with that of \citet{Peacock10}. The photometric offsets and rms scatters of the differences between their measurements and our new magnitudes are: $\Delta U = 0.010\pm0.016$ with $\sigma=0.222$ (192 clusters in common), $\Delta B = 0.010\pm0.010$ with $\sigma=0.170$ (296 clusters in common), $\Delta V = 0.071\pm0.008$ with $\sigma=0.133$ (295 clusters in common), $\Delta R = 0.289\pm0.007$ with $\sigma=0.126$ (292 clusters in common), and $\Delta I = 0.203\pm0.011$ with $\sigma=0.178$ (277 clusters in common) (this study minus Peacock et al. 2010).}
\end{figure*}

\citet{bh00} presented self-consistent, optical ($UBVRI$) photometric data for 285 M31 GCs. \citet{gall04} updated the Bologna Catalog \citep{batt87} with the homogenized optical ($UBVRI$) photometry collected from many previous catalogs available in the literature, which is named the Revised Bologna Catalog (RBC). In order to obtain optical photometric data as homogeneous as possible, these authors took as a photometric reference the data by \citet{bh00}. The RBC has been updated collecting the latest data for M31 star clusters and cluster candidates. The latest RBC (V.5) used here was updated on 2012 August. \citet{fan10} performed $UBVRI$ photometry for 970 M31 star clusters and cluster candidates based on archival images of the LGGS. \citet{Peacock10} identified M31 star clusters using the images from the UK Infrared Telescope's Wide Field Camera (WFCAM) and the Sloan Digital Sky Survey (SDSS) archives, and presented photometry for 945 M31 star clusters and cluster candidates in the SDSS $ugriz$ and the near-infrared (NIR) $K$ bands.

Figures 7--10 show the comparison of our photometry of the clusters considered here with previous photometric data in the RBC, with previous photometric measurements in \citet{bh00}, \citet{fan10} and \citet{Peacock10}. In Figure 10, we have transformed the $ugriz$ magnitudes to the Johnson-Cousins $UBVRI$ magnitudes using the equations given by \citet{Jester05}.

From Figures 7--10, we can see that our new magnitudes in the $U$, $B$ and $V$ bands are in good agreement with those in previous studies, while ours in the $R$ and $I$ bands are systematically fainter than those of previous studies. However, the photometric scatters in the $R$ and $I$ bands are the same as those in the other three bands. So, we consider that the photometric offsets in the $R$ and $I$ bands result from the formulas (4) and (5). If we add an appropriate constant value in the formulas (4) and (5), the photometric offsets between this study and previous studies in the $R$ and $I$ bands will disappear. In addition, there are some star clusters, of which the magnitudes in the $UBVRI$ bands between this study and previous studies (Galleti et al. 2004; Barmby et al. 2000; Fan et al. 2010; Peacock et al. 2010) are larger than 1.0 mag, i.e., our aperture magnitudes are fainter than previous measurements in \citet{gall04}, \citet{bh00}, \citet{fan10}, and \citet{Peacock10}. This is probably due to the very bright and uneven background around the clusters. These clusters are B129, B254, B333, B391-G328, DAO55, and NB16. We reported their names in Figures 7--10 (black circles).

\section{MASSES OF M31 OLD STAR CLUSTERS}

\citet{cald11} estimated the masses of the sample star clusters based on the assumption of $M/L_{\rm V}=2$. However, the results of \citet{str09,str11} showed that M31 star clusters have the values of $M/L_{\rm V}$ between 0.27 and 4.05. In this paper, we will estimate the masses of the sample clusters by comparing their homogenous photometry obtained here with stellar population synthesis models. We use the {\sc galev} SSP models \citep[e.g.,][]{kurth99,schulz02,anders03} for our estimates as \citet{wang12} used. In fact, the masses of star clusters obtained here are independent of SSP models adopted here (see details in Section 3.2). The {\sc galev} models provide absolute magnitudes (in the Vega system) in 77 bands for SSPs of $10^6~M_\odot$, including Johnson $UBVRI$ \citep{lan83}. The difference between the intrinsic absolute magnitudes and those given by the model provides a direct measurement of the cluster mass, in units of $10^6~M_\odot$. To reduce mass uncertainties resulting from photometric uncertainties based on only magnitudes in one band (in general, the $V$ band is used), we estimate the masses of the sample star clusters using the magnitudes in the $UBVRI$ bands obtained here. The masses of clusters obtained based on the magnitudes in different bands are a little different, therefore we averaged them as the final cluster mass which may better reflect the true mass of a star cluster. The magnitudes of the sample clusters in the $UBVRI$ bands can be obtained using the equations (1)--(5). In order to correctly estimate their masses, we will consider the offsets of magnitudes between ours and those of previous studies. However, the captions of Figures 6--9 showed that these offsets are different. So, as \citet{gall04} did, we take as a photometric reference the dataset by \citet{bh00}, i.e., our new photometry has been transformed to this reference list by applying the offsets we derived here. In our estimates, the masses of the sample clusters are not independent of their ages and metallicities, which were taken from \citet{cald11}. In addition, the reddening values were taken from \citet{kang12}, who derived reddening values from three methods: (1) mean reddening values from available literature \citep{bh00,fan08,cald09,cald11}; (2) median reddening values of star clusters located within an annulus at each 2 kpc radius from the center of M31 (for these star clusters there are not available reddening values in the literature); and (3) for star clusters at distances larger than 22 kpc from the center of M31, the foreground reddening values of $E(B-V)=0.13$ are adopted. We considered that the reddening values from \citet{kang12} are more reasonable. In fact, the reddening values of \citet{kang12} are generally in agreement with those of the literature \citep{bh00,fan08,cald09,cald11}. We cannot estimate masses of 7 sample clusters, since \citet{cald11} did not present their ages and metallicities. The resulting mass estimates of 297 old star clusters are listed in Table 2 with their $1\sigma$ uncertainties. We also list the absolute magnitudes in the $V$ band ($M_V$) and the reddening values ($E(B-V)$) used here.

\begin{figure*}
\centerline{
\includegraphics[scale=0.65,angle=-90]{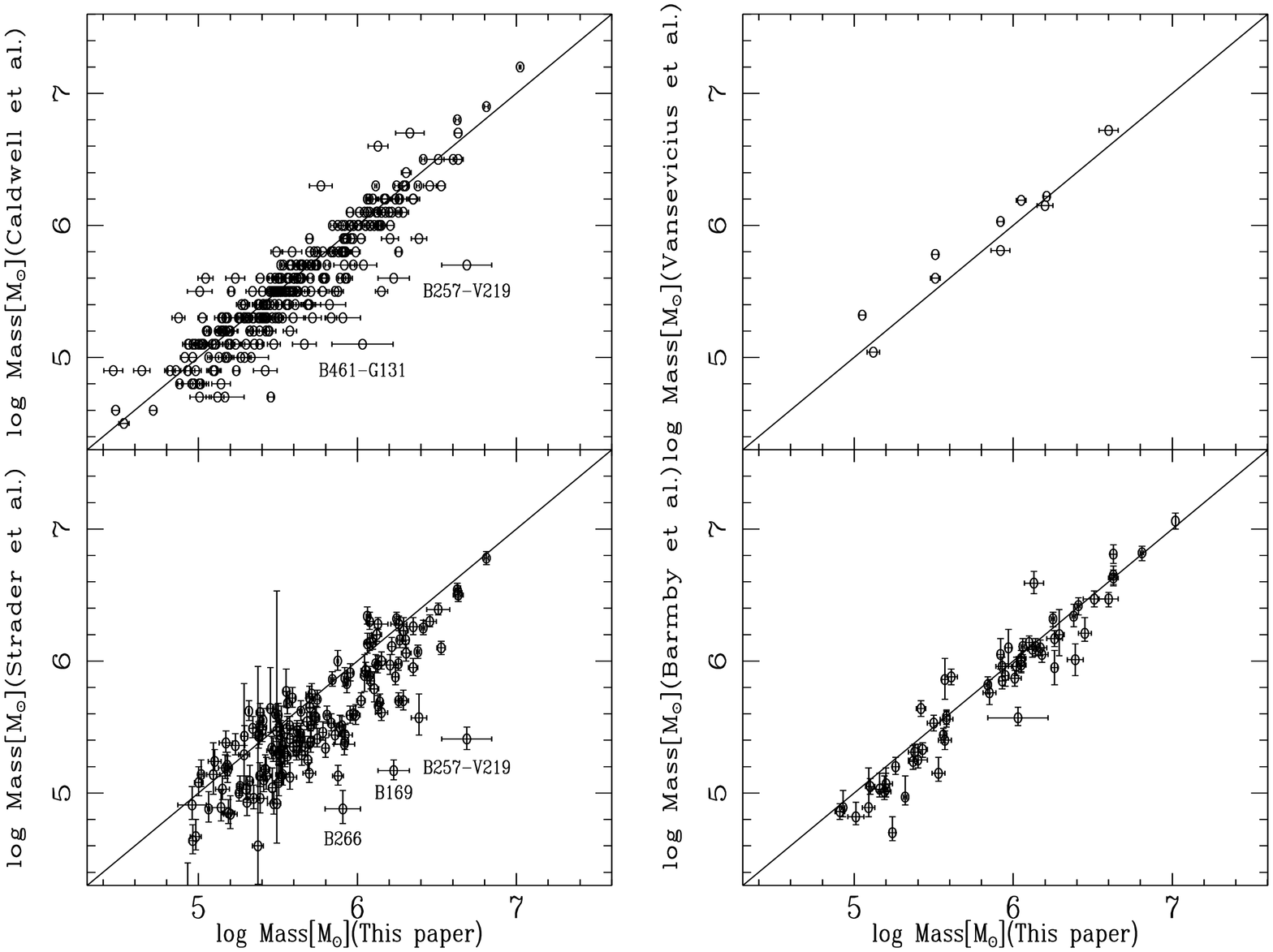}}
\caption[]{Comparison of the masses obtained here with those published by \citet{barmby07,barmby09}, \citet{vanse09}, \citet{cald11}, and \citet{str11}. The mass offsets and rms scatters of the differences between their measurements and ours are: $\Delta {\rm log~Mass} = 0.055\pm0.012$ with $\sigma=0.202$ (297 clusters in common) (this study minus Caldwell et al. 2011); $\Delta {\rm log~Mass} = -0.077\pm0.042$ with $\sigma=0.134$ (10 clusters in common) (this study minus Vansevi{\v c}ius et al. 2009); $\Delta {\rm log~ Mass} = 0.202\pm0.021$ with $\sigma=0.262$ (156 clusters in common) (this study minus Strader et al. 2011); and $\Delta {\rm log~Mass} = 0.065\pm0.022$ with $\sigma=0.170$ (60 clusters in common) (this study minus Barmby et al. 2007, 2009).}
\label{fig10}
\end{figure*}

\subsection{Comparison with Previous Masses}

Figure 11 shows the comparison of our masses with those of previous studies (Barmby et al. 2007, 2009; Vansevi{\v c}ius et al. 2009; Caldwell et al. 2011; Strader et al. 2011). In previous studies,  \citet{barmby07,barmby09}, \citet{vanse09}, and \citet{cald11} estimated masses from the mass-to-light ratio ($M/L$) coupled with estimated ages, while \citet{str11} determined masses based on the velocity dispersions and structural parameters of star clusters. In general, our estimated masses are in agreement with those published by \citet{cald11}, \citet{vanse09}, and \citet{barmby07,barmby09}, while ours are slightly larger than those of \citet{str11}. In addition, it is interesting that, for the most massive star cluster in M31, B037-V327, our new mass is in good agreement with those published by \citet{cald11} and \citet{barmby07,barmby09}. There are two star clusters, of which the mass scatters (${\rm log~Mass}$) between this study and \citet{cald11} are large, 0.93 for B461-G131 and 0.99 for B257-V129, respectively. We reported their names in Figure 11 (black circles). The large mass scatters result from the adoption of different reddening values: for B257-V219, 0.66 by \citet{cald11} and 1.17 here, respectively; for B461-G131, 0.10 by \citet{cald11} and 0.52 here, respectively. If we adopt the reddening values of \citet{cald11}, our estimated masses of these two clusters will be in good agreement with those of \citet{cald11}. There are three star clusters, of which the mass scatters (${\rm log~Mass}$) between this study and \citet{str11} are larger than 1.0. We also reported their names in Figure 11 (black circles). Since \citet{str11} determined their masses based on a dynamical model, we cannot find out what causes the large mass scatters for these three clusters. Other works \citep{Beasley04,perina10,kang12} also determined the masses of M31 star clusters. Three clusters of \citet{perina10} are included in our sample. \citet{perina10} estimated the masses $({\rm log~Mass})>4.7$ for B083-G146 and B347-G154, and $({\rm log~Mass})>4.8$ for NB16, which are in agreement with ours: 5.42 for B083-G146, 5.61 for B347-G154, and 5.21 for NB16. The sample star clusters here are not included in \citet{Beasley04} and \citet{kang12}.

\begin{figure*}
\centerline{
\includegraphics[scale=0.65,angle=-90]{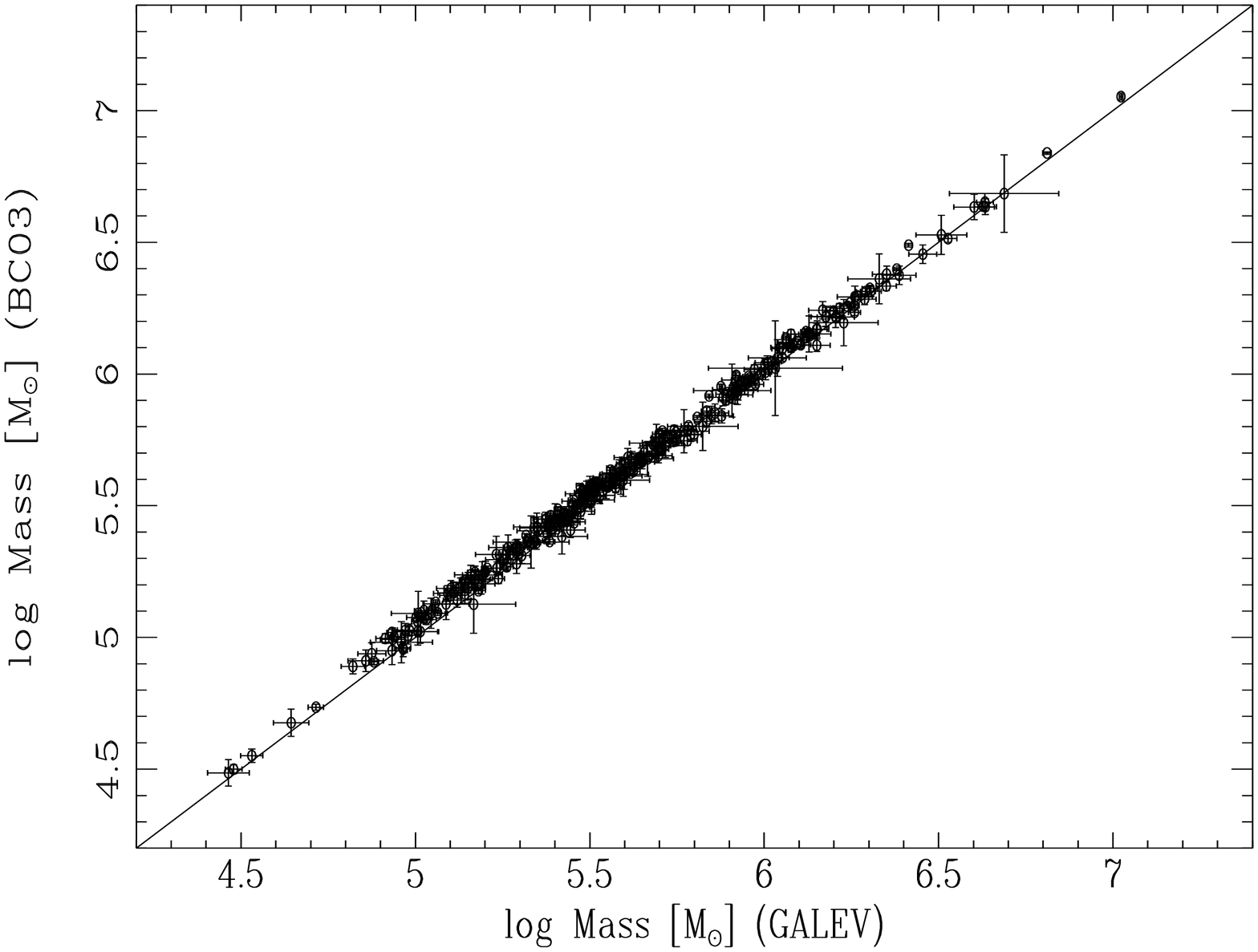}}
\caption[]{Comparison of the masses obtained based on different SSP models: BC03 versus {\sc galev}. The mass offsets and rms scatters of the differences between BC03 model and {\sc galev} model are: $\Delta {\rm log~Mass} = 0.032\pm0.002$ with $\sigma=0.028$ (297 clusters in common).}
\end{figure*}

\begin{figure*}
\centerline{
\includegraphics[scale=0.65,angle=-90]{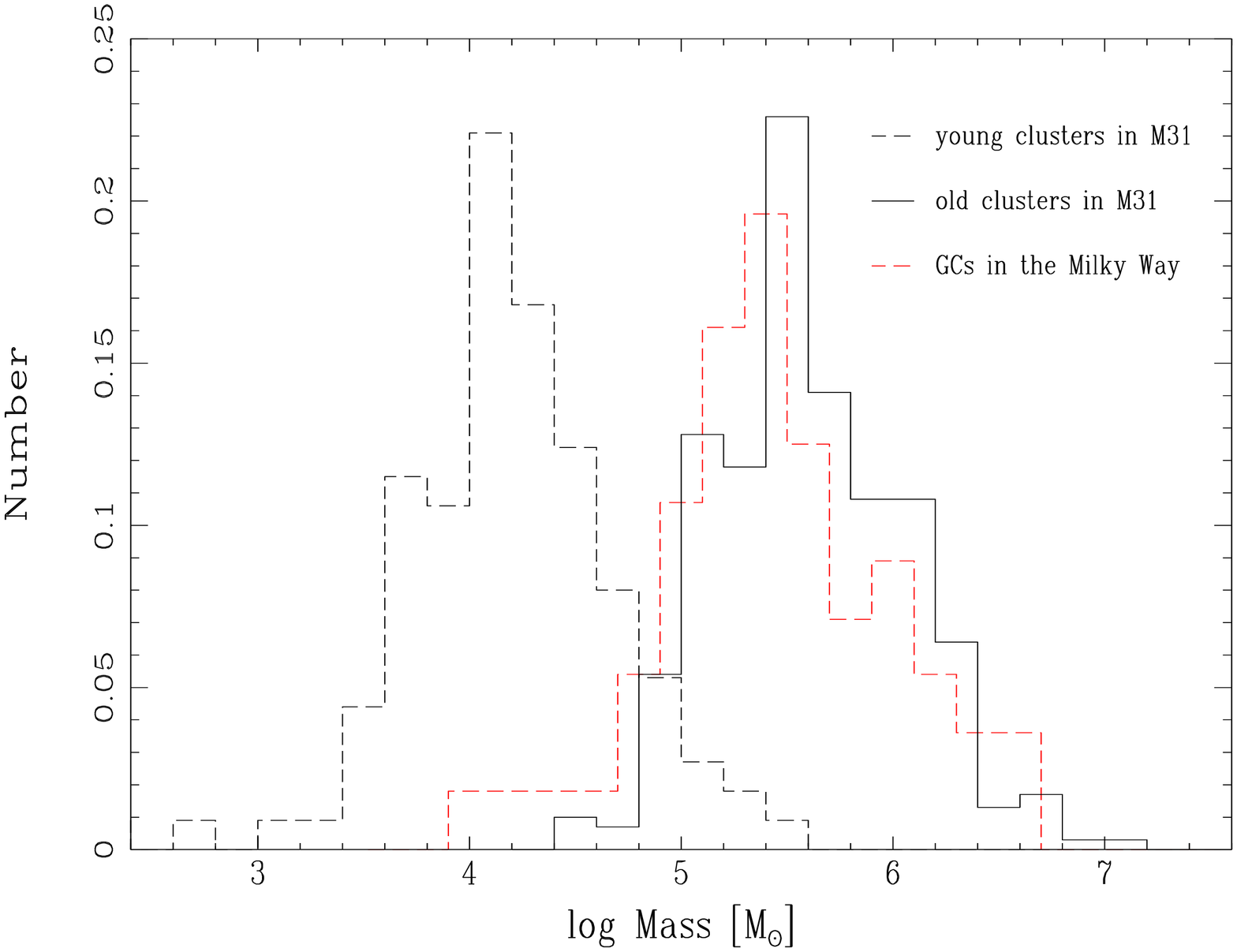}}
\caption{Mass distribution of young, old star clusters in M31, and GCs in the Milky Way.}
\end{figure*}

\subsection{Comparison of Masses between Different SSP Models}

In order to investigate the differences of star cluster masses caused by adopting different SSP models, we also calculated the masses of the sample star clusters using the SSP models of \citet{bc03} (hereafter BC03). BC03 models are normalized to a total mass of $1 M_\odot$ in stars at age $t=0$. The absolute magnitudes in $UBVRI$ bands are included in the BC03 SSP models. The difference between the intrinsic absolute magnitudes and those given by the model provides a direct measurement of a cluster mass. Figure 12 presented the comparison of masses estimated using the {\sc galev} and BC03 models, which showed that the masses obtained using different SSP models are consistent.

\section{DISCUSSION OF OLD STAR CLUSTERS}

\subsection{Distribution of Star Cluster Masses}

Figure 13 presents the distribution of masses of the sample clusters obtained here. The distributions of masses of the young star clusters in M31 \citep{wang12} and the GCs in the Milky Way \citep{pryor93} are also shown for comparison. It is shown that the masses of M31 old star clusters extend from $\sim 3\times10^4 M_\odot$ to $\sim 10^7 M_\odot$, with the peak of $\sim 4\times10^5 M_\odot$. For the young star clusters in M31, the masses extend from $\sim 5\times10^2 M_\odot$ to $\sim 2\times10^5 M_\odot$, with the peak of $\sim 3\times10^4 M_\odot$. It is obvious that some young star clusters are more massive than some old star clusters. From Figure 13, we also note that, for the GCs in the Milky Way, the masses extend from $\sim 10^4 M_\odot$ to $\sim 5\times10^6 M_\odot$, with the peak of $\sim 3\times10^5 M_\odot$, indicating that the massive GCs in M31 are more than those in the Milky Way.

\subsection{Correlation between Mass and Luminosity}

\begin{figure*}
\centerline{
\includegraphics[scale=0.65,angle=-90]{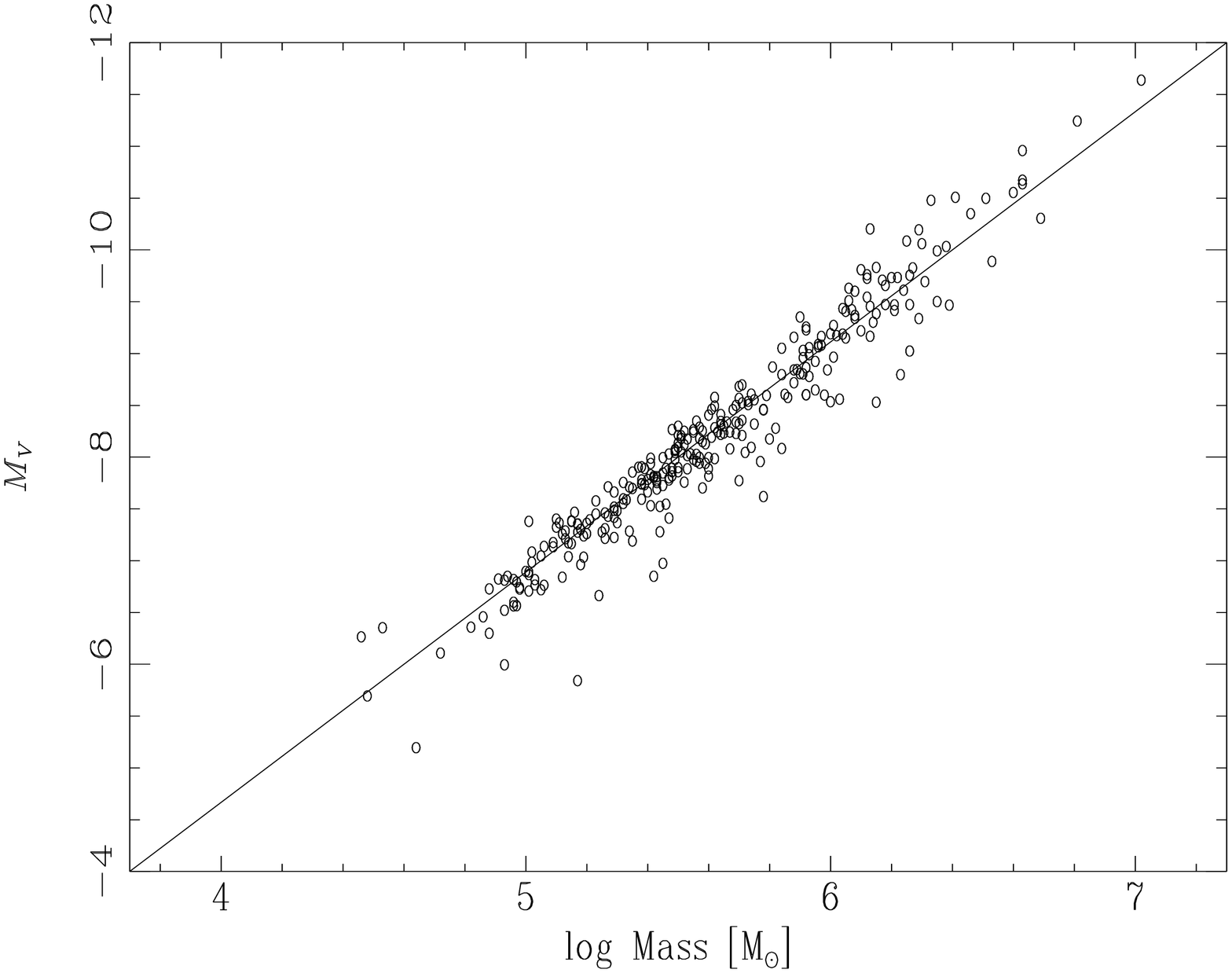}}
\caption{Correlation between Mass and Luminosity for old star clusters in M31. $\Delta ({\rm 2.5log~Mass}+M_V)= 5.764\pm0.063$ with $\sigma=1.081$ (297 clusters in common).}
\end{figure*}

In order to estimate photometric masses of star clusters, $M/L_V$ is generally needed to be assumed. In \citet{cald11}, the $M/L_V = 2$ is assumed independent of [Fe/H]. However, stellar population models predict that $M/L_V$ should be dependent on age and metallicity \citep[see Figure 2 of][for details]{{mclau08}}.
Figure 14 shows the correlation between the masses and the absolute magnitudes for M31 old star clusters, both of which are obtained here. The scatter superimposed on the correlation may indicate equivalently the variations of the mass-to-light ratio from cluster to cluster.

\begin{figure*}
\centerline{
\includegraphics[scale=0.65,angle=-90]{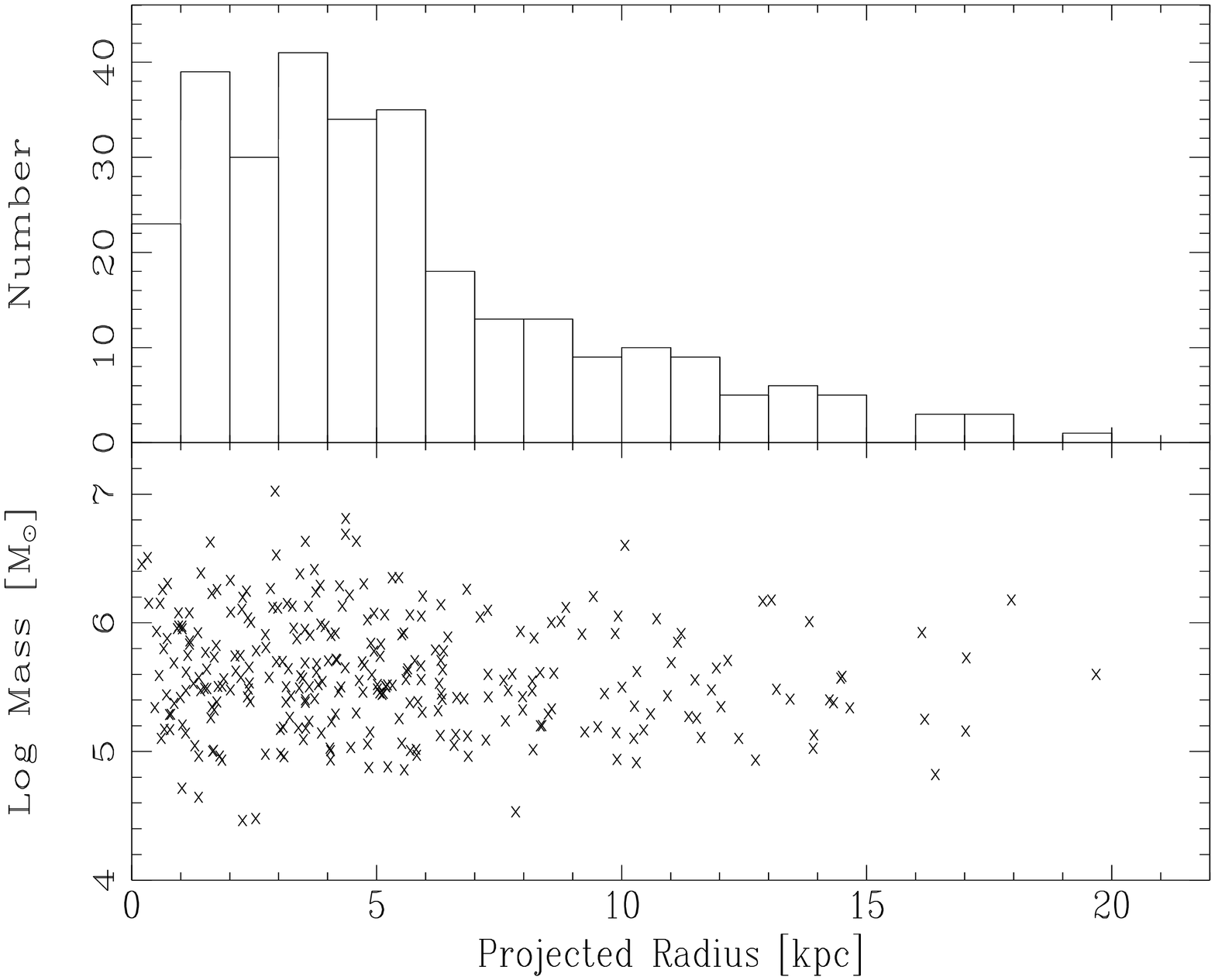}}
\caption{Top: number histogram of sample old star clusters against projected radius. Bottom panel: mass vs. projected radius for sample old star clusters.}
\end{figure*}

\subsection{Spatial Distribution}

Figure 15 shows the number, and masses of old star clusters as a function of projected radius ($R_{\rm gc}$) from the center of M31. We adopted a central position for M31 at $\rm\alpha_0=00^h42^m44^s.30$ and $\rm \delta_0=+41^o16'09''.0$ (J2000.0) following \citet{hbk91} and \citet{per02}. Formally,
\begin{equation}
X=A\sin\theta+B\cos\theta,
\end{equation}

\begin{equation}
Y=-A\cos\theta+B\sin\theta,
\end{equation}
where $A=\sin(\alpha-\alpha_0)\cos\delta$ and $B=\sin\delta\cos\delta_0 - \cos(\alpha-\alpha_0) \cos\delta \sin\delta_0$. We adopt a position angle of $\theta=38^\circ$ for the major axis of M31 \citep{ken89}. In the top panel, the histogram for the radial distribution of old star clusters shows clearly most of the sample clusters exist in the central region of $R_{\rm gc}<6 {\rm kpc}$. The bottom panel presents that there is weak evidence that massive clusters like to be at small $R_{\rm gc}$, implying that either strong dynamical friction drives predominantly more massive GCs inward, or massive GCs may favor forming in the nuclear regions of galaxies with higher pressure and density \citep{Georgiev09}.

\section{SUMMARY}

The main results of this paper are:

(1) We presented photometry in 15 intermediate-band filters for 304 old star clusters in the field of M31 based on the images observed by the BATC Multicolor Sky Survey.

(2) We estimated the masses of these sample star clusters by comparing between the homogenous photometry obtained here with the SSP models.

(3) The masses of these old star clusters have masses between $\sim 3\times10^4 M_\odot$  and $\sim 10^7 M_\odot$.

(4) The peaks of young and old star clusters are $\sim 3\times10^4 M_\odot$ and $\sim 4\times10^5 M_\odot$, respectively.

\acknowledgments
We would like to thank the anonymous referee for providing rapid and thoughtful report that helped improve the original manuscript greatly. This study has been supported by the Chinese National Natural Science Foundation through grants 11373035, 11433005, 11373033, 11203034, 11203031, 11303038, 11303043, 11073032, 11003021, and 11173016, and by the National Basic Research Program of China (973 Program), No. 2014CB845702, 2014CB845704, 2013CB834902, and by the joint fund of Astronomy of the National Nature Science Foundation of China and the Chinese Academy of Science, under Grants U1231113.

\clearpage
\begin{table*}
\small \setlength{\tabcolsep}{0.3em} \caption{BATC Photometry of 304 M31 Old Star Clusters} \vspace{0mm} \label{t1.tab}
\begin{tabular}{ccccccccccccccccc}
\tableline\tableline
ID & $a$ & $b$ & $c$ & $d$ & $e$ & $f$ & $g$ & $h$ & $i$ & $j$ & $k$ & $m$ & $n$ & $o$ & $p$ & $r_{\rm ap}$\\
 & (mag) & (mag) & (mag) & (mag) & (mag) & (mag) & (mag) & (mag) & (mag) & (mag) & (mag) & (mag) & (mag) & (mag) & (mag) & (\arcs)\\
\hline
 AU010          &...     &...     &18.930  &18.182  &17.812  &17.687  &17.531  &17.371  &17.098  &17.047  &17.013  &17.056  &17.030  &16.839  &16.730  &  3.34\\
                &...     &...     &0.232   &0.219   &0.218   &0.233   &0.284   &0.280   &0.281   &0.285   &0.329   &0.355   &0.385   &0.374   &0.382   &      \\
 B001-G039      &...     &18.723  &18.186  &17.804  &17.533  &17.193  &16.786  &16.632  &16.315  &16.230  &16.100  &15.935  &15.764  &15.678  &15.607  & 10.02\\
                &...     &0.112   &0.033   &0.093   &0.078   &0.031   &0.035   &0.021   &0.015   &0.023   &0.023   &0.020   &0.034   &0.044   &0.038   &      \\
 B002-G043      &...     &18.665  &18.237  &17.972  &17.881  &17.763  &17.483  &17.425  &17.319  &17.289  &17.220  &17.215  &17.098  &17.013  &17.192  &  5.01\\
                &...     &0.056   &0.022   &0.057   &0.049   &0.028   &0.033   &0.020   &0.017   &0.032   &0.031   &0.034   &0.061   &0.068   &0.105   &      \\
 B003-G045      &...     &18.837  &18.399  &18.250  &18.149  &17.797  &17.531  &17.376  &17.228  &17.139  &17.065  &17.015  &16.823  &16.752  &16.652  &  6.68\\
                &...     &0.083   &0.027   &0.087   &0.079   &0.036   &0.040   &0.026   &0.020   &0.035   &0.037   &0.032   &0.060   &0.062   &0.078   &      \\
 B004-G050      &19.752  &18.409  &17.968  &17.666  &17.282  &17.131  &16.792  &16.680  &16.545  &16.417  &16.322  &16.268  &16.163  &16.077  &16.017  &  6.68\\
                &0.540   &0.059   &0.019   &0.052   &0.038   &0.021   &0.020   &0.015   &0.011   &0.020   &0.019   &0.017   &0.035   &0.039   &0.048   &      \\
 B005-G052      &17.919  &17.360  &16.805  &16.119  &16.038  &15.819  &15.497  &15.344  &15.130  &15.056  &14.926  &14.869  &14.772  &14.695  &14.565  &  8.35\\
                &0.058   &0.029   &0.026   &0.026   &0.020   &0.013   &0.014   &0.011   &0.010   &0.011   &0.015   &0.010   &0.015   &0.021   &0.024   &      \\
 B006-G058      &18.165  &17.219  &16.585  &16.124  &15.790  &15.620  &15.285  &15.199  &15.005  &14.878  &14.747  &14.697  &14.593  &14.522  &14.426  & 11.69\\
                &0.073   &0.041   &0.007   &0.008   &0.005   &0.004   &0.005   &0.005   &0.005   &0.005   &0.007   &0.005   &0.012   &0.008   &0.014   &      \\
 B008-G060      &19.255  &18.454  &17.870  &17.410  &17.061  &16.891  &16.535  &16.409  &16.231  &16.061  &15.943  &15.863  &15.740  &15.769  &15.640  &  8.35\\
                &0.124   &0.078   &0.015   &0.019   &0.013   &0.012   &0.012   &0.012   &0.012   &0.015   &0.017   &0.014   &0.026   &0.022   &0.033   &      \\
 B009-G061      &19.022  &17.947  &17.600  &17.350  &17.191  &16.941  &16.776  &16.718  &16.503  &16.438  &16.453  &16.343  &16.417  &16.196  &16.347  & 10.02\\
                &0.110   &0.067   &0.030   &0.036   &0.034   &0.031   &0.035   &0.035   &0.037   &0.037   &0.049   &0.044   &0.062   &0.049   &0.089   &      \\
 B010-G062      &18.897  &18.280  &17.536  &17.213  &16.992  &16.816  &16.534  &16.439  &16.233  &16.164  &16.088  &15.990  &15.875  &15.833  &15.786  & 10.02\\
                &0.102   &0.073   &0.012   &0.017   &0.013   &0.010   &0.012   &0.011   &0.011   &0.015   &0.019   &0.014   &0.027   &0.024   &0.038   &      \\
\hline
\end{tabular}
\end{table*}
{$^1$}{This table is available in its entirely in machine-readable.}

\clearpage
\begin{table*}
\small \setlength{\tabcolsep}{0.3em} \caption{Masses of 297 M31 Old Star Clusters} \vspace{0mm} \label{t2.tab}
\begin{tabular}{ccccccccc}
\tableline\tableline
ID &  log Mass ($U$) & log Mass ($B$) & log Mass ($V$)  & log Mass ($R$)  & log Mass ($I$)  & log Mass (average) & $M_V$  &  $E(B-V)$ \\
 & $(M_\odot)$ & $(M_\odot)$ & $(M_\odot)$ & $(M_\odot)$ & $(M_\odot)$ & $(M_\odot)$
& (mag) & (mag)\\
\hline
AU010      &     ...      &$5.18\pm0.13$ &$5.12\pm0.18$ &$5.11\pm0.11$ &$4.97\pm0.37$ &$5.10\pm0.04$ &$ -7.470$&0.21 \\
B001-G039  &     ...      &$5.61\pm0.05$ &$5.60\pm0.02$ &$5.61\pm0.01$ &$5.64\pm0.04$ &$5.61\pm0.01$ &$ -8.260$&0.25 \\
B002-G043  &     ...      &$5.00\pm0.03$ &$4.95\pm0.02$ &$4.90\pm0.01$ &$4.99\pm0.07$ &$4.96\pm0.02$ &$ -6.887$&0.01 \\
B003-G045  &     ...      &$5.17\pm0.04$ &$5.12\pm0.02$ &$5.10\pm0.01$ &$5.14\pm0.06$ &$5.13\pm0.01$ &$ -7.280$&0.16 \\
B004-G050  &$5.48\pm0.19$ &$5.48\pm0.03$ &$5.47\pm0.01$ &$5.40\pm0.01$ &$5.41\pm0.04$ &$5.45\pm0.02$ &$ -7.915$&0.13 \\
B005-G052  &$6.34\pm0.04$ &$6.29\pm0.01$ &$6.15\pm0.01$ &$6.12\pm0.01$ &$6.07\pm0.02$ &$6.21\pm0.05$ &$ -9.483$&0.22 \\
B006-G058  &$6.18\pm0.05$ &$6.17\pm0.01$ &$6.15\pm0.01$ &$6.10\pm0.01$ &$6.10\pm0.01$ &$6.14\pm0.02$ &$ -9.370$&0.11 \\
B008-G060  &$5.69\pm0.09$ &$5.66\pm0.01$ &$5.65\pm0.01$ &$5.60\pm0.01$ &$5.61\pm0.03$ &$5.64\pm0.02$ &$ -8.287$&0.17 \\
B009-G061  &$5.50\pm0.07$ &$5.47\pm0.02$ &$5.41\pm0.02$ &$5.36\pm0.01$ &$5.36\pm0.06$ &$5.42\pm0.03$ &$ -7.882$&0.09 \\
B010-G062  &$5.56\pm0.08$ &$5.60\pm0.01$ &$5.56\pm0.01$ &$5.53\pm0.01$ &$5.56\pm0.03$ &$5.56\pm0.01$ &$ -8.416$&0.20 \\
\hline
\end{tabular}
\end{table*}
{$^2$}{This table is available in its entirely in machine-readable.}
\end{document}